\documentclass[12pt]{iopart}

\usepackage{harvard}
\usepackage{graphicx}
\usepackage{subcaption}

\begin{document}
\bibliographystyle{dcu}
\citationmode{abbr}

\title[Beams Selection In Non-Coplanar Radiotherapy]{On-line Dose Calculation Using Deep Learning for Beams Selection in Non-Coplanar Radiotherapy}


\author{Fang GUO\textsuperscript{1}, Franklin OKOLI\textsuperscript{1}, Ulrike Schick\textsuperscript{1,2}, Dimitris Visvikis\textsuperscript{1}\textsuperscript{*}, Antoine Valeri\textsuperscript{1,2}, Julien Bert\textsuperscript{1,2}}

\address{\textsuperscript{1} LaTIM, INSERM UMR1101, Brest, France}
\address{\textsuperscript{2} Brest University Hospital, France}
\address{\textsuperscript{*} Author to whom any correspondence should be addressed.}

\ead{dimitris.visvikis@univ-brest.fr}
\vspace{10pt}
\begin{indented}
\item[]March 2023
\end{indented}

\begin{abstract}
Non-coplanar Intensity-Modulated Radiation Therapy (IMRT) goes a step further by orienting the gantry carrying the radiation beam and the patient couch in a non-coplanar manner to accurately target the cancer region and better avoid organs-at-risk. The use of a non-coplanar treatment trajectory significantly enhances the degree of freedom and flexibility but increases drastically the complexity of the optimization.
In inverse planning optimization the dose contribution for all potential beam directions is usually pre-calculates and pre-loads into the Treatment Planning System (TPS). The size the dose matrix becomes more critical when moving from coplanar IMRT to non-coplanar IMRT since the number of beams increases drastically.
A solution would be to calculate “on-the-fly” the dose contribution to each new candidate beam during optimization. This is only possible if a dose calculation engine is fast enough to be used online during optimization iterations, which is not the case in standard method.
Therefore, in this work we propose an IMRT optimization scheme using deep learning based dose engine to compute the dose matrix on-line. The proposed deep learning approach will be combined into a simulated-annealing-based optimization method for non-coplanar IMRT.  Since the dose engine will compute the dose contribution on-line during the optimization, the final main optimization method requires to keep in memory a very lightweight dose matrix.
The proposed method was compared with clinical data showing a good agreement considering dosimetry of the treatment plans. The main advantage of the proposed method was the reduction of the memory storage from 9GB to 10MB during the optimization process.
\end{abstract}

%
%
%
%
%

\section{Introduction}

Non-coplanar Intensity-Modulated Radiation Therapy (IMRT) goes a step further by orienting the gantry carrying the radiation beam and the patient couch in a non-coplanar manner to accurately target the cancer region and better avoid organs-at-risk \cite{bortfeld2006imrt}. Despite improvements achieved from using non-coplanar IMRT, they are still not adopted for use at most treatment centers. The use of a non-coplanar treatment trajectory significantly enhances the degree of freedom and flexibility but increases drastically the complexity of the optimization \cite{Smyth2019}.

In inverse planning optimization the dose contribution of each beam is precalculated and stored into a dose influence matrix with a huge size. However, considering memory storage and optimization time, the matrix dimension has to be small-scale so it can be used clinically. Common methods limit the number of beams, i.e., searching space of the optimal beam orientations, by increasing the sampling angle’s interval. 
Similarly, the number of voxels that describe the patient is also limited by resampling the CT image into large voxel spacing. Even fast dose algorithm, such collapsed cone dose algorithm \cite{hasenbalg2007collapsed} and the superposition-convolution algorithm \cite{jenkins2012monte} for examples, not allows to compute the dose contribution of a given in real time during the optimisation loop. Therefore, the dose influence matrix, is usually pre-calculates and pre-loads into the Treatment Planning System (TPS) for all potential beam directions. The size the dose influence matrix becomes more critical when moving from coplanar IMRT to non-coplanar IMRT since the number of beams increases drastically. Such a strategy is usually considered time-consuming and storage-occupying \cite{jelen2005finite}.

One solution would be to further reduce the number of voxels and the number of possible beams candidate to obtain a reasonable size of the dose influence matrix. However, the full benefit of the non-coplanar could not be demonstrated because the quality of the treatment plan would deteriorate. Another solution would be to calculate the dose contribution to each new candidate beam during optimization. This corresponds to no longer pre-calculating the entire influence matrix but only the useful values "on the fly". This is only possible if a dose calculation engine is fast enough to be used online during optimization iterations.

Recently, many researchers have introduced Artificial Intelligence (AI), especially deep learning technology, into medical applications. Deep learning is a class of machine learning algorithms that uses multiple layers to extract higher-level features from the raw input progressively. There are several medical deep-learning applications for dose prediction that are proven to be successful. 

There is numerous research work that is able to predict dosemap in a very fast manner \cite{nguyen2018three,nguyen2019feasibility,fan2019automatic}. A first attempt to use an AI-based dose engine in a TPS was proposed in \cite{liu2021neuraldao}. However, the work was limited to the direct aperture optimization (DAO) for IMRT, meaning that the beam selection was not considered.  But the aim was the same, trye to reduce the dose influence matrix.

In this work we propose to go further by considering a deep learning based dose engine that depend of the  beam configurations, and then brings the work of \cite{liu2021neuraldao} to a higher level, by considering beam angle selection as well. The proposed deep learning approach will be combined into a simulated-annealing-based optimization method for non-coplanar IMRT.  Since the dose engine will compute the dose contribution on-line during the optimization, the final main optimization method requires to keep in memory a very lightweight influence matrix. 

\section{Materials and methods}

The deep learning approach to predict the dose is first presented in the following sections. This work was similar to previous work from \cite{nguyen2018three,nguyen2019feasibility,fan2019automatic}. However, for a fast execution the inputs parameters, like the beam angulation, were directly injected in the lattence space, which was different to the state-of-the art method. Then a dedicated section will described how the optimization algorithm was combined with the proposed dose engine. Finally an evaluation study will be presented.

\subsection{Training dataset }
To fully train our network, we need adequate dosemaps under different beam directions and the corresponding 3D CT images. To create such a dataset, we use an open-source program package called \textit{matRad} \cite{cisternas2015matrad,craft2014shared}, which is a Matlab package for radiotherapy. We generated a dataset consisting of 40 patients with head and neck cancer. The CT images are retrieved from the Head-Neck-Radiomics-HN1 dataset \cite{test} in the Cancer Image Archive (TCIA). We randomly split the first 30 patients’ data into a training set and validation set (24 patients for the training dataset and 6 patients for the validation dataset), and the rest 10 patients form a testing set to evaluate the performance of the network and the whole algorithm. The original size of CT images were resized to $64\times64\times 64$ voxels. We use matRad to estimate the dose distribution. For a given beam angle the dose influence matrix is calculated from matRad and the MLC optimisation is solved using DAO method \cite{unkelbach2015optimization} and the planning target volume (PTV). This PTV was provided  as an 3D image binary mask, and will be also used as an input of the deep learning method. The final 3D dose map from the colimated beam was used into the trainning data set. The size of the dose was the same that the CT image, $64\times64\times 64$ voxels.  The sampling angle was $5^\circ$ for both the gantry and couch angle, leading to 169 possible beam directions. Therefore for the 40 patients, we obtained 6760 sets in total. For each of this set we computed the corresponding dose map. All images and angles were normalized from 0 to 1. The final trainning data set was composed of the 3D CT image, the gantry and couch pair angle, the PTV mask, and the corresponding 3D dosemap.

\subsection{Network architecture}
\begin{figure}
	\centering    
	\begin{subfigure}{1.0\textwidth}
		\includegraphics[width=1.0\linewidth]{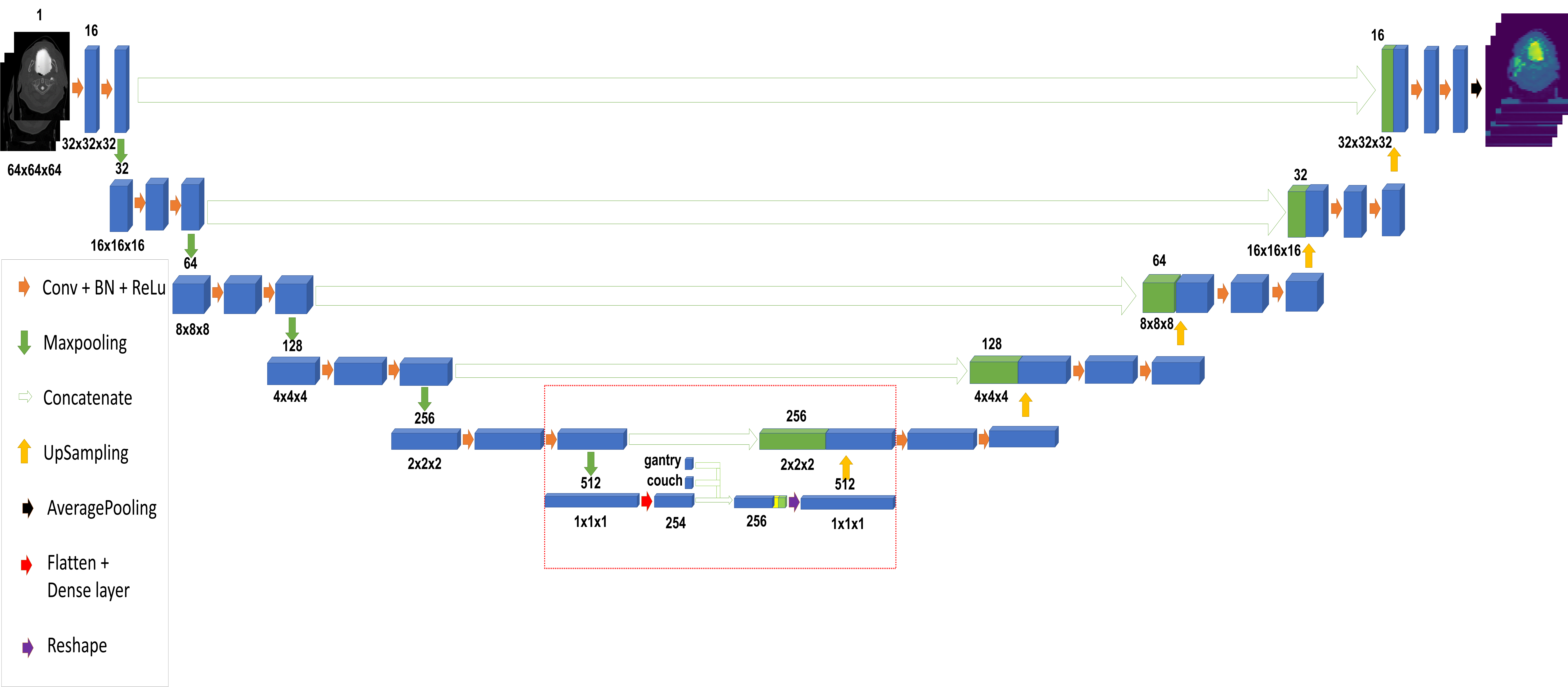}
	\end{subfigure}
	\hfill
	\hfill  
	\caption{Architecture of the proposed network where the bottleneck part was modified to consider the gantry and couch inputs.}     
	\label{fig:Pred-Unet}    
\end{figure}

The network for the dose prediction was based on a 3D U-Net architecture  \cite{ronneberger2015u}, which is wildly applied in medical applications. This 3D U-Net model, as shown in \Fref{fig:Pred-Unet}, uses the 3D CT images, the value of gantry angle, and couch angles as inputs and the PTV mask. The output is the predicted 3D dosemap with a size of $64\times64\times64$ voxels, which correspond to the dose distribution of the shaped beam according patient anatomy from the CT and the PTV.  CT image and PTV mask have the same size and were fed to the network as two input channels. 

Gantry and couch angles are encoded uing only 2 parameters, while CT and PTV are encoded with almost half a million due to the number of voxels. Then, if we introduce angles parameters directly as standard input at the beginning of the network, these two angles will be not considered by the network compare to the other image parameter, due to the numerous of downsamplings and flatting. A solution consist encoding the beam angles into a 3D image with a same dimension that CT and PTV. This is mostly achieved by using raytracing approach which consist to draw in 3D the beam oriented and shaped in an image. However, this apporach increase the total time to recover a dose map, since a pre-processing step of raytracing is required for each new beam orientation.

Therefore, the standard architecture was slightly modified to encode the couch and ganrty angle directly in the latent space through the bottleneck part. The normalized couch and gantry angles were concatenated with the one-dimensional array of the bottleneck and then passed to a fully connected layer to let all neurons receive the beam angle information. After this fully connected layer, the result is reshaped and passed to the right part of the U-Net. In every layer, the activation function is ReLU (Rectified Linear Unit) in order to merge the information on the beam angle into the network. The network architecture was developed and trained on Keras, using the TensorFlow backend.

\subsection{Network training}

 
 In the training phase, we use the adaptive momentum algorithm (ADAM) \cite{kingma2014adam} for minimizing the loss function with a learning rate of $1\times10^{-4}$, the loss function is a mean-square error function defined as follow, 
 
 \begin{equation}
 	MSE=\frac{1}{n}\sum_{i=1}^{n}(Y_i - \hat{Y}_i)^2
 	\label{eq: loss}
 \end{equation}
 
 where $i$ is the pixel index, $n$ total number of pixels, $Y_i$ is the ground truth image and $\hat{Y}_i$ is the predicted image. During the network training process, both training and validation loss was monitored. Consequently, the training process was stopped at the epoch when the validation loss ceased showing improvement with respect to training loss, and reach aproximatively 400 epochs. To balance training efficiency and GPU capacity, in every training epoch, the batch size was 8. The model was trained on an NVIDIA Corporation GP106GL [Quadro P2000] with 6GB of dedicated RAM. 


\subsection{Non-coplanar IMRT optimization algorithm}

Avoiding local minimum may be a difficulty when there are a large number of optimization variables and constraints. Simulated Annealing (SA) is a probabilistic technique for approximating the global optimum of a given objective function, by simulating a cooling process. This method is a meta-heuristic algorithm that allows solve problem optimization with ample search space. The key point of the SA method is that when the temperature is high, the algorithm is exploring solution, by discarding even candidate that improve the cost function, in order not to fall into a local minimum. This temperature is cooling down iteratively to slowy beginning accepting solution according a probability calculated by the amount of improvement that bring the considered solution, a random factor and the current temperature. At the end when the temperature is low, the method converge to a solution considered as an approximation of the global optimum.

\begin{figure}
	\centering
	\includegraphics[width=0.5\linewidth]{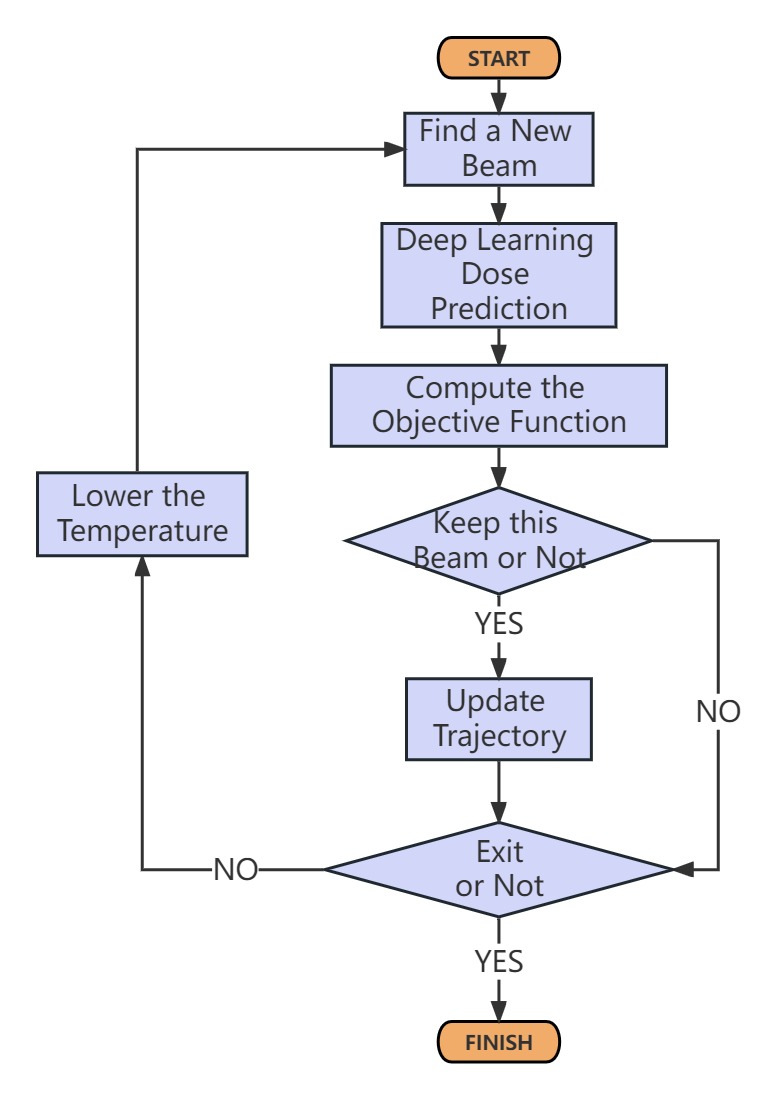}
	\caption{Scheme that summarize the proposed SA optimization method with dose engine from deep learning approach}     
	\label{fig:SA-DDL}    
\end{figure}

This optimisation method was already and successfully used in numerous treatment planning optimisation \cite{villa_fast_2022}. We propose an optimization scheme that allows to combine the proposed dose engine and the SA method to optimize a non-coplanar IMRT treatment. This optimization algorithm scheme, named SA-DDL (Simulated Annealing Dose Deep Learning), is described in \Fref{fig:SA-DDL}. 

The algorithm starts with a high temperature $\mathbf{T}_{init}$, and an empty beam candidate set (formed of control points) $\mathbf{B_K}$. A the begining of each iteration a new solution representing by beam candidates $\mathbf{B_{K_{new}}}$ is proposed by updating the current set $\mathbf{B_K}$. This update is a random process that decide to add new random control point or replace or remove one of the beam candidate randomly. Subsequently, the proposed AI-based dose engine is used to estimate the dose distribution according to the new solution  $\mathbf{B_{K_{new}}}$. Direct Aperture Optimization is not apply since it performed within the dose prediction based on the PTV mask. AI prediction implicitely shaped the beam. Then the computation of the objective function is performed. According to this new score the simulated annealing method determines whether or not to accept this solution. If accepted the beam candidates set $\mathbf{B_{K_{new}}}$ is swap this the current set $\mathbf{B_K}$. If not accepted the current set $\mathbf{B_K}$ is kept as it is. Finally, the temperature of the SA method is cooling down according a cooling factor, and a new iteration start. The optimization process stop when severals conditions are reached, such as maximum number of iterations, minimal temperature or a cost function that reach a plateau.

The objective function is defined in \Eref{eq:obj_fun}, where $V_{PTV}$ denotes the PTV set and $V_{OAR}$ denotes the Organ At Risk (OAR) set. A penalty factor $p^+$ controlls the relative importance of the target tumor voxels and $p^-$ controlls the relative importance of organ-at-risk voxels. The variable ${d^P}$ denotes the prescribed dose and $d$ the dose computed from the predicted dosemap. The idea of the objective function is to minimize the least-square deviation between the prescribed dose and the actual dose received by the tumor voxels. 

\begin{equation}
	f(d) = \frac{1}{N_t}\sum_{j\in V_{PTV}}p^+(d_j-{d^P}_j)^2 + \frac{1}{N_o}\sum_{j\in V_{OAR}}p^-(d_j-{d^P}_j)^2
	\label{eq:obj_fun}
\end{equation}

\section{Evaluation Study}

\subsection{AI-based dose engine validation}
We evaluate first the dose engine provided by the deep learning method. We then compute the predicted dose map for the $10$ patients from the testing dataset and compare the results with the dose map from matRad dose engine. Since each patient is treated with different beam directions, for each of these patients we calculated the dose map for severals beams. In order to estimate dose maps from realistic beam direction, for each patient a treatment plan was calculated using matRad. Then for each patient $169$ beam directions was considered. The network performance was mainly assessed using the mean dose and relative dose difference ratio for the PTV and OARs. There are defined as follows:
 
\begin{enumerate}    
	\item \textbf{Mean Dose} Mean of the dose absorbed by all voxels in a specific region of interest to the patient. Given $\mathbf{N}$ voxels in a region of interest $\mathbf{V}$ with each voxel $j$ receiving dose $d_j$, it is calculated using the relation:
	\begin{equation}
		Dose_{Mean} = \frac{\sum_{j=1}^{N}}{d_j}\forall \in \mathbf{V}
	\end{equation}
	
	\item \textbf{Relative Dose Difference Ratio} Relative difference of the radiation dose between two dose values $Dose_1$ and $Dose_2$:
	\begin{equation}
		Relative\ Difference\ Ratio = |\frac{Dose_1 - Dose_2}{Dose_1}| \times \%
	\end{equation}
\end{enumerate}


\subsection{Non-coplanar Treatment Planning}
The aim of this evaluation was to show that the dose engine from AI based approach not change the treatment plan to a standard dose engine, here the one provided by matRad. The purpose of using AI-based dose engine was not to improve the quality of the treatment plan but to reduce the optimisation time and storage needs.

This was evaluated by using the 10 patient from the test set. For each of this patient a treatmen plan for non-coplanar IMRT treatment was optimized using SA and the dose engine from matRad (SA-matRad) and using SA and the dose deep learning engine (SA-DDL). The same parameters, prescribed dose, OAR dose constraints, PTV, etc. were used for both optimizations. In addition to the mean dose and the relative dose difference ratio, Dose Volume Histogram (DVH) were also estimated.


\subsection{Computing time and memory storage}

For both optimization methods, computing time and the memory storage was monitored. The computing time was measure for each iteration. The memory storage measure the data size need to optimize the beam selection, especially the dose influence matrix but also the data size of the deep learning model.


\section{Results}

\subsection{AI-based dose engine}
\begin{figure}
	\centering
	\includegraphics[width=0.6\linewidth, height=0.55\linewidth]{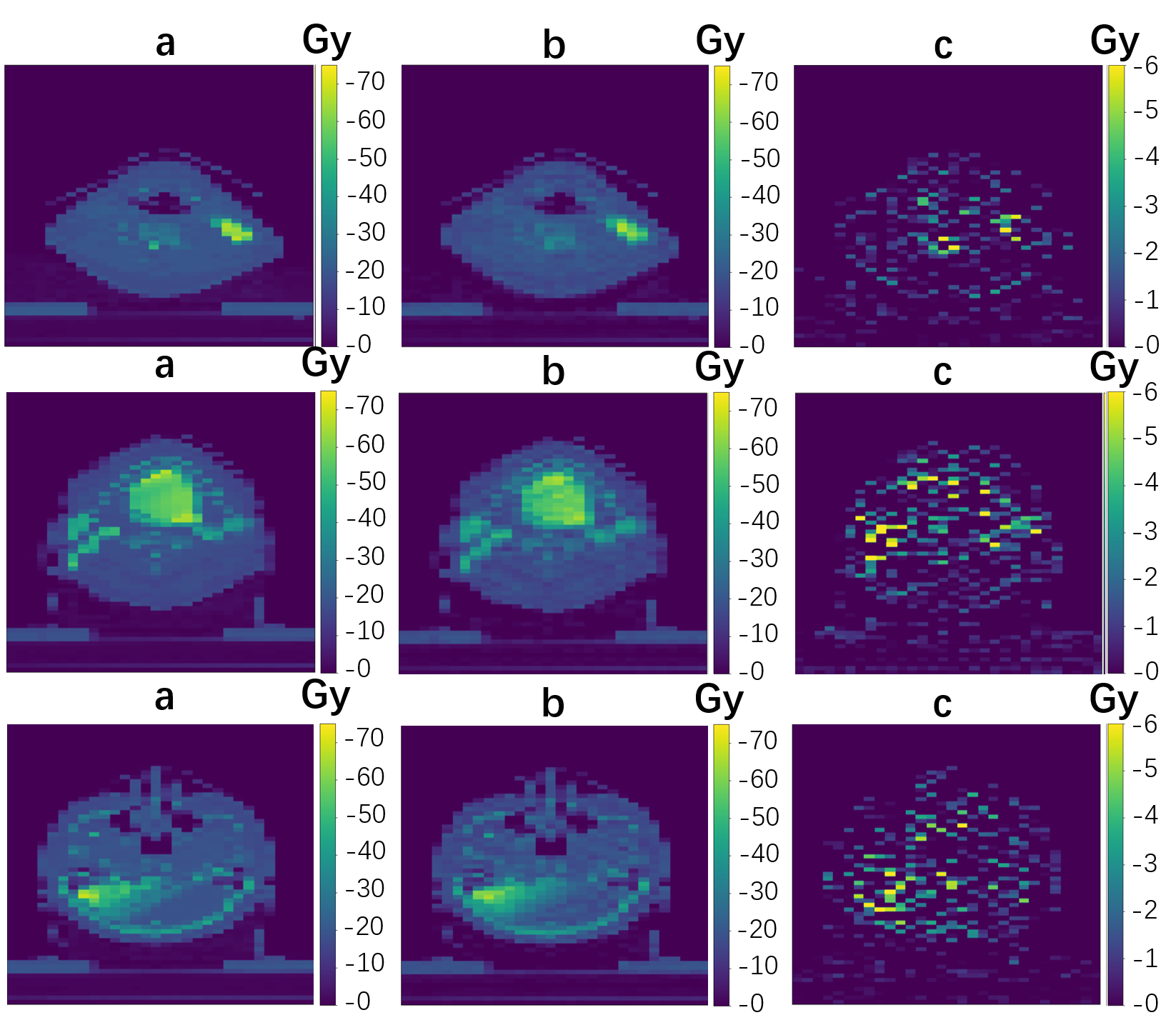}
	\caption{Dosemap comparison example \textbf{(a)} dosemaps generated in matRad \textbf{(b)} predicted dosemaps \textbf{(c)} absolute difference between \textbf{(a)} and \textbf{(b)}. Each row represents a different 2D slice of the patient head.}
	\label{fig:DM} 
\end{figure}

Table \ref{fig:DM} shows the comparison between the matRad planned dosemap and the predicted dosemap. Column (a) is the dosemap generated by matRad, column (b) is the predicted dosemap, and column (c) is the difference between the first two columns. 

\begin{figure}
	\centering    
	\includegraphics[width=0.75\linewidth]{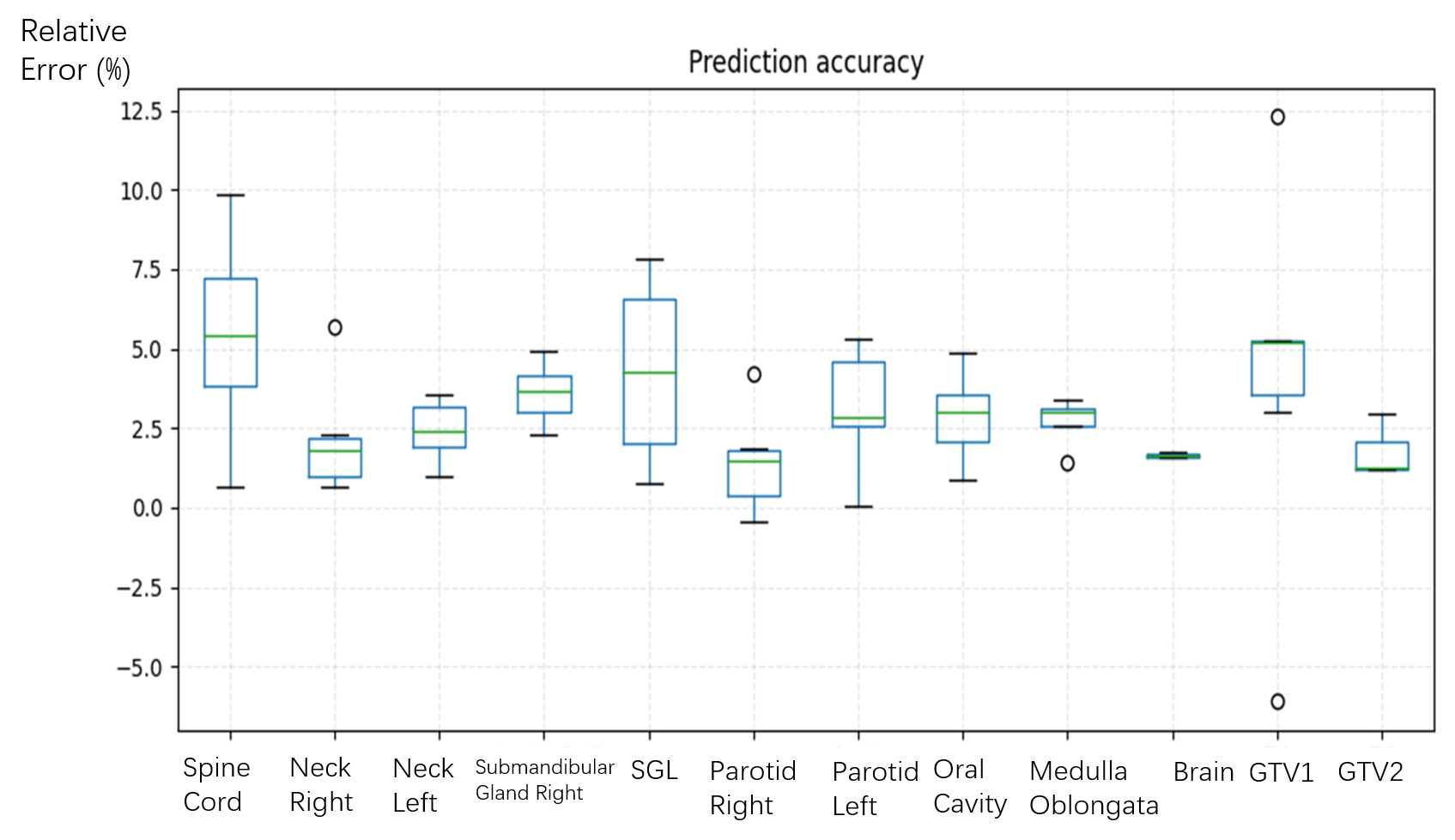}
	\caption{Prediction accuracy for 10 patient test sets. Relative error between dose from prediction (dashed lines) and matRad dose engine (solid lines) were given for each target volume and organ at risks.} 
	\label{fig:Boxplot_net}    
\end{figure}

\begin{figure}
	\centering    
	\begin{subfigure}[b]{0.49\textwidth}
		\centering
		\includegraphics[width=1.0\linewidth]{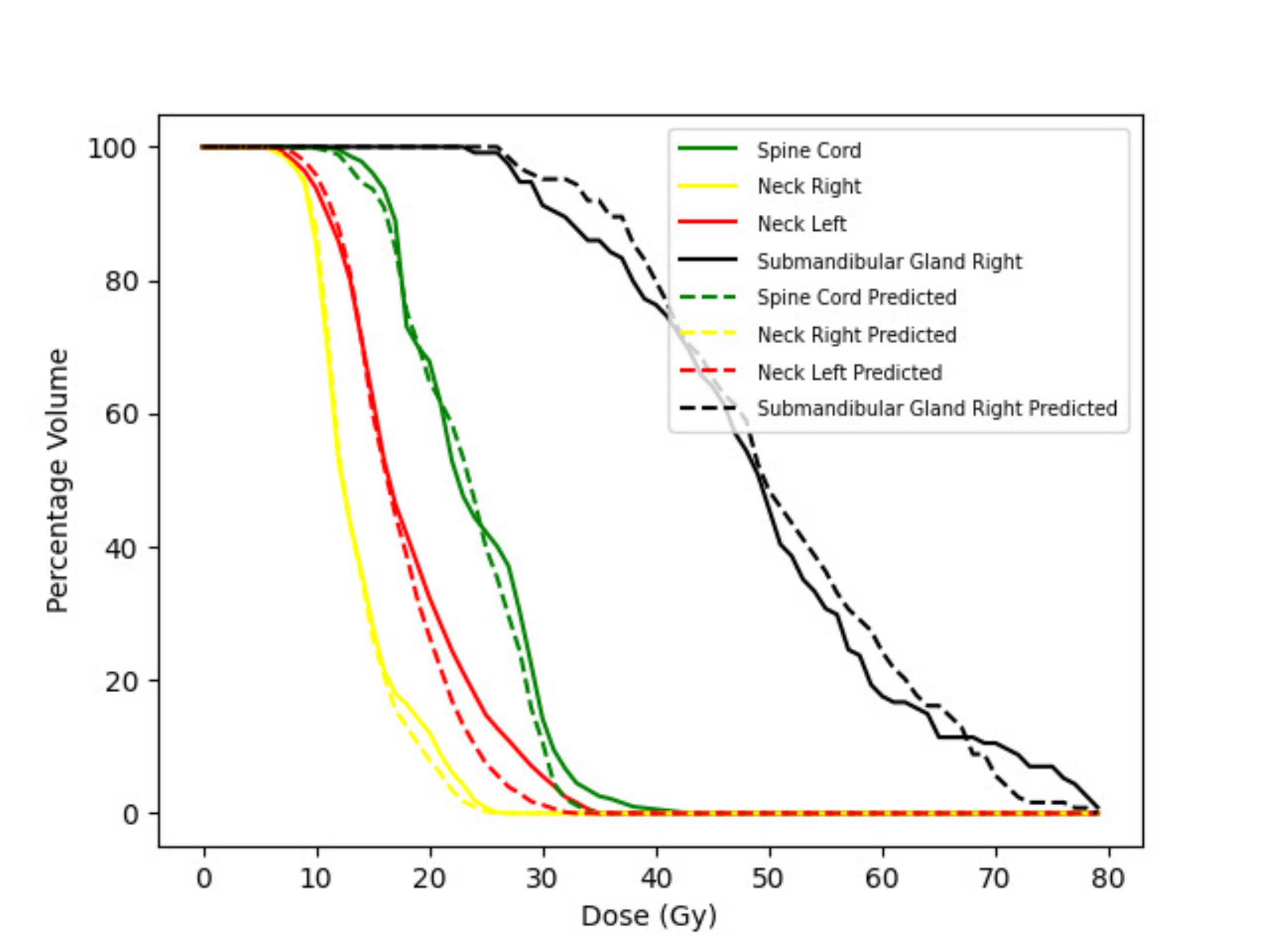}
		\caption{ }
	\end{subfigure}
	\begin{subfigure}[b]{0.49\textwidth}
		\centering
		\includegraphics[width=1.0\linewidth]{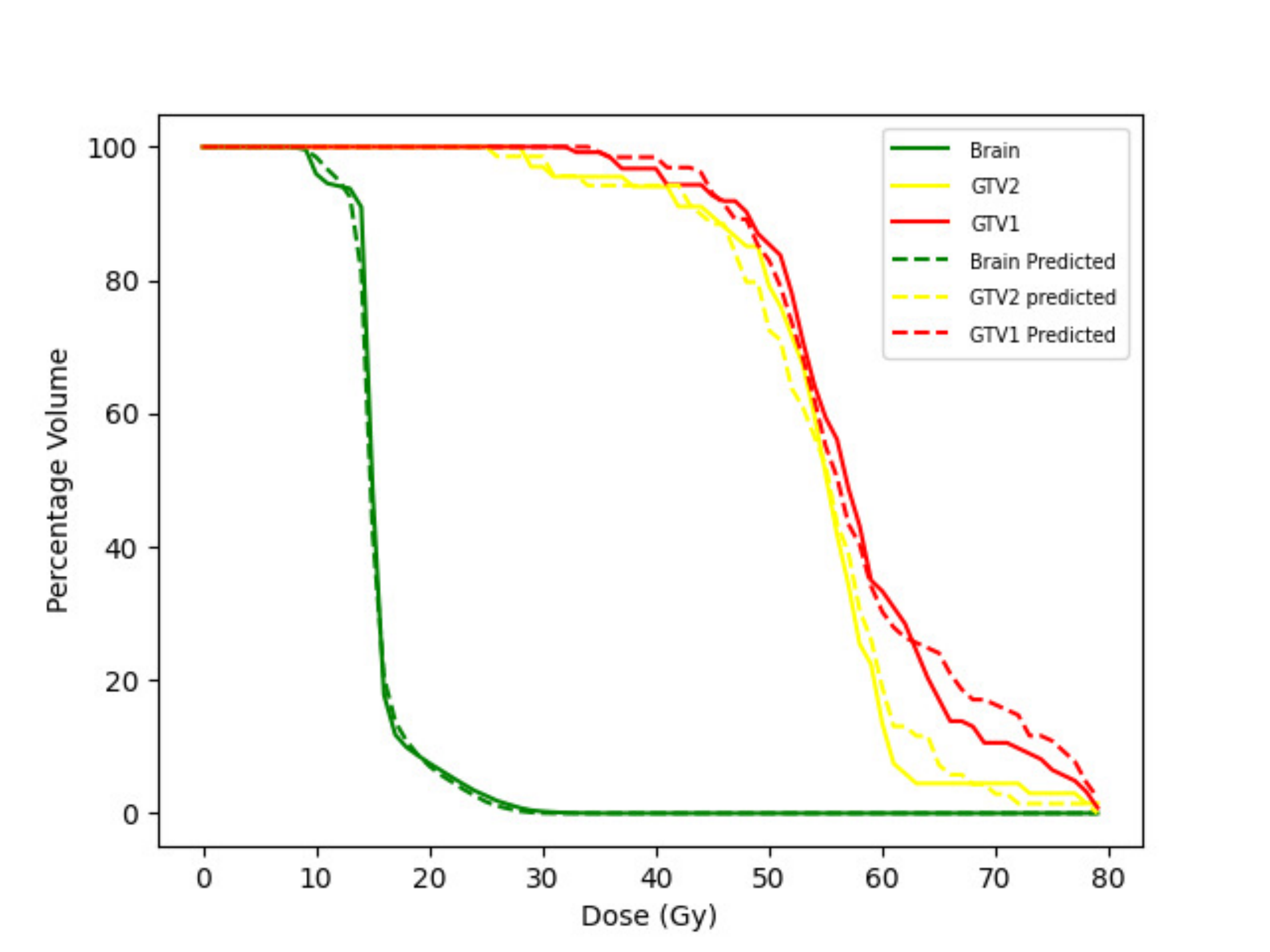}
		\caption{ }
	\end{subfigure}
	\hfill  
	\caption{DVH comparison between dose map from deep learning prediction and the dose engine from matRad for an arbitrary patient within the test for a $gantry=180^\circ, couch=180^\circ$. (a) and (b) represents the different OARs and PTVs.}     
	\label{fig:DVH_pred}    
\end{figure}

Prediction accuracy for 10 patient test sets and for the target volumes and organs at risks are show in \Fref{fig:Boxplot_net}. 
The result have shown that the average error between the most OARs and PTVs are in average close to $5\pm 3\%$. For example in patient 1, the brain shows a mean relative error of $1.75\pm0.49\%$, while the spinal cord which occupies few voxels yields the value of  $9.85 \pm 2.37\%$. This average error is higher than the other PTVs and OARs, this is explained by the fact that the spinal cord's area in CT images is significantly small compared to other tissues, thus one small difference in pixel value may result in a noticeable influence on dose computation. DVH for an arbitrary selected patient wthin the test data set were plotted in \Fref{fig:DVH_pred} for the dose map obtained by the deep learning approach and the dose engine provided by matRad. Results have shown a good agreement between the DVH. All of these results validate the suitability of thte deep learning solution to estimate dose map with an accuracy enough for treatment planning optimisation.

\subsection{Non-coplanar Treatment Planning}

\Tref{tab:tps} present the result of dose comparison between treatment plan obtained by the simulated annelaing approach with deep learning dose engine (SA-DDL) and the one obtained by the seimulated annealing and matRad dose engine. These restuls are given in average for the 10 patient within the test data set. We can see that there are no obvious signs stating that one method is better than the other regarding dose, for GTV. 

\begin{figure}
	\centering    
	\begin{subfigure}[b]{0.7\textwidth}
		\centering
		\includegraphics[width=1.0\linewidth]{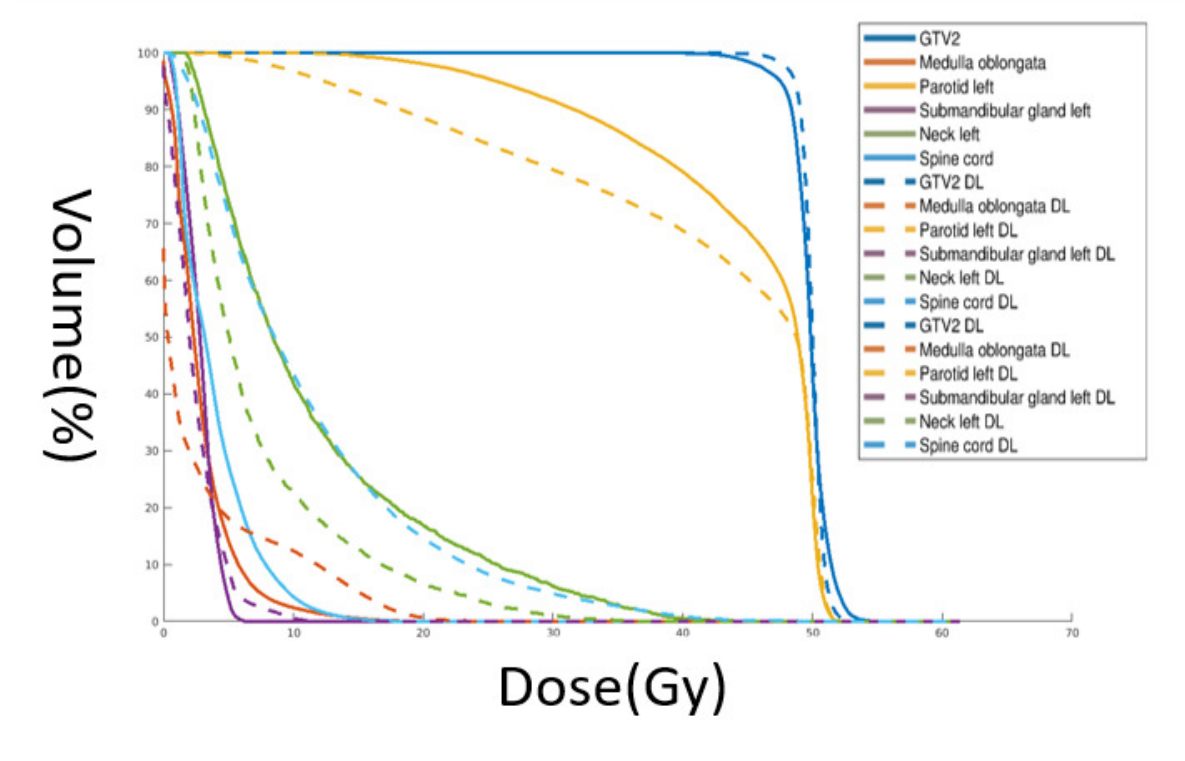}
		\caption{ }
	\end{subfigure}
	\hfill
	\begin{subfigure}[b]{0.7\textwidth}
		\centering
		\includegraphics[width=1.0\linewidth]{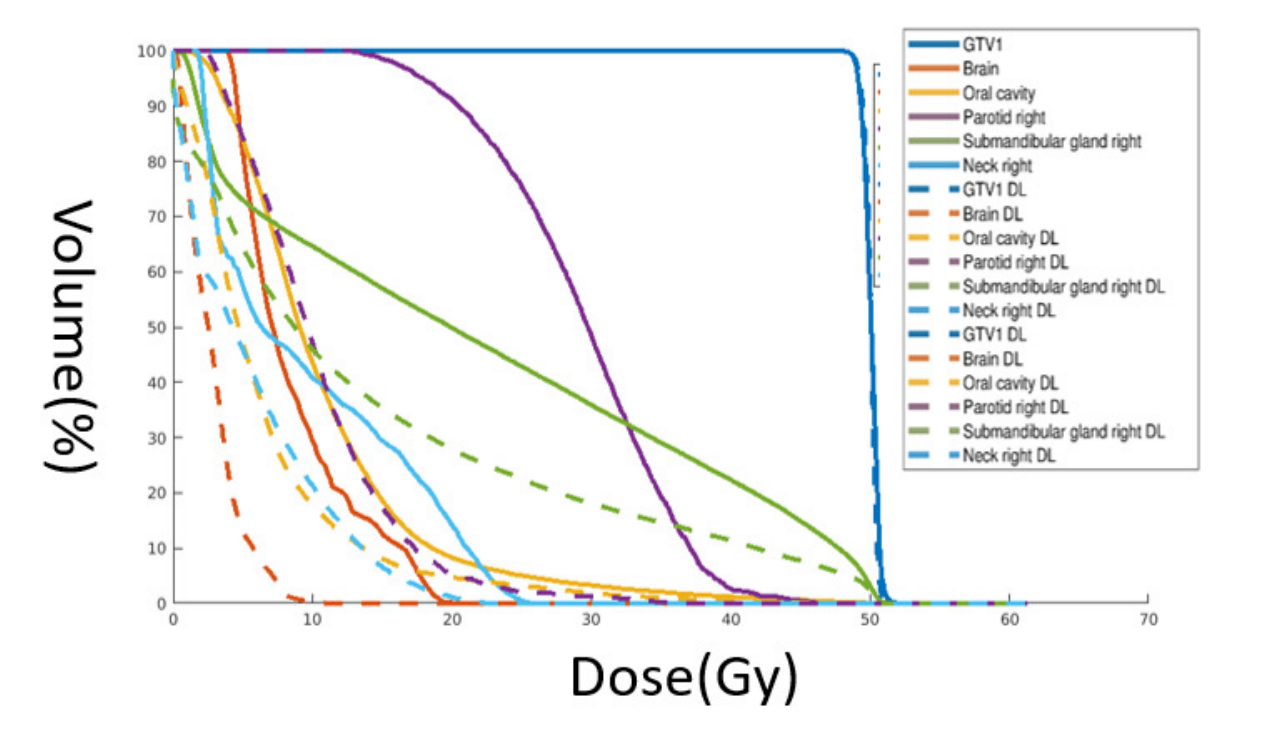}
		\caption{ }
	\end{subfigure}
	\hfill  
	\caption{DVH comparison between dose map from treatment plan calculated using SA-matRad (solid lines) and SA-DDL (dashed lines)}     
	\label{fig:simu_DL_DVH}    
\end{figure}

DVH for both dose maps from the two treatment plan are shown in \Fref{fig:simu_DL_DVH}. Doses from the SA-DDL were slighlty lower but except for spinal cord and brain. Therefore we can observed that the proposed method using deep learning dose engine performed at least at the same level of quality and can be considered as equivalent. Of course extend evaluation with more clinical data should be investigated to confirm this first preliminary observation.

As shown in \Tref{tab:time} the main advantage of the proposed method is the use of small amoung of memory storage 10MB for the optimization of non-coplanar treatment plan. Standard method that use pre-calculation dose matrix apporach, like SA-matRad need 9GB of memory storage. Since no memory access is large data size is required with deep learning approach, and since GPU-based prediciton is fast, SA-DDL was faster with 6s per iteration compare to SA-matRad with 180s.


\begin{table}
	\caption{Average relative dose difference between non-coplanar treatmant plan from SA-matRad and SA-DDL for the 10 patient test set.}
	\label{tab:tps}
	\begin{indented}
		\item[]\begin{tabular}{@{}lll}
			\br
			\textbf{OARs $\&$ PTVs}	& \textbf{Relative Mean Dose}	& \textbf{STD}\\
			\mr
			Spinal Cord		& $-6.86\%$			& $0.283$\\
			Neck Right		& $12.00\%$			& $0.184$\\
			Neck Left       & $11.96\%$         & $0.155$\\
			Submandibular Gland Right       & $2.35\%$         & $0.089$\\
			Submandibular Gland Left        & $-0.89\%$        & $0.261$\\
			Parotid Right   & $7.65\%$          & $0.227$\\
			Parotid Left    & $2.92\%$          & $0.301$\\
			Oral Cavity     & $4.07\%$          & $0.165$\\
			Medulla Oblongata               & $-2.01\%$        & $0.255$\\
			Brain           & $3.71\%$          & $0.313$\\
			GTV2            & $-3.54\%$         & $0.079$\\
			GTV1            & $3.04\%$          & $0.097$\\
			\br
		\end{tabular}\\
	\end{indented}
\end{table}

\begin{table}
	\caption{Comparison of predicition time ans memory storage between SA-matRad and SA-DDL.}
	\label{tab:time}
	\begin{indented}
		\item[]\begin{tabular}{@{}lll}
			\br
			\textbf{Methods}	& \textbf{Iteration time}	& \textbf{Data size}\\
			\mr
			SA-matRad (pre-calculated dose matrix)	& 180s			& 9GB\\
			SA-DDL (on-line dose calculation)	& 6s			& 10 MB\\
			\br
		\end{tabular}\\
	\end{indented}
\end{table}

\section{Discussion}
In this manuscript, we developed a new optimization algorithm SA-DDL for non-coplanar IMRT treatment planning. This algorithm uses the simulated annealing optimization method to find the optimal beam directions, when computing the dose cost at different control points we use a pre-trained neural network to predict the dose distribution.

The performance of the deep learning approach has been evaluated. The average dose difference for most OARs and PTVs was around $5\%$, and the standard deviation close to $3\%$, which is acceptable for fast treatment planning optimisation. The DVH shows that the predicted dosemap was close to the matRad generated dosemap. Therefore, we can conclude that our method is reliable and is able to predict the dosemap according to different beam configurations. 

Comparison between treatment plan from SA-matRad and the proposed method SA-DDL were studied. Results have mainly shown equivalence in term of dosimetry. Which was the main objectif of the study in oder to proof that the method that reduce memory storage by using on-line dose calculation was not introducing approximation within the optimization. However, both plan are not easy to interprate since some of the OAR are highers from the SA-matRad method compare to SA-DDL and some are lowers. A large sutdy with more clinical data should be explored in order to confirm if both are statistically equivalent.

Any way, the aim of this paper was a proof of concept to show that deep learning dose engine may be used to reduce drastically the memory usage, from 9GB to 10MB, during the treatment planning optimisation in non-coplanar beam radiotherapy.

 \section*{Acknowledgement}
This work was partially supported by Brest métropole océane.

\section*{References}

\bibliography{refs.bib}

@article{Smyth2019,
	author = {Smyth, Gregory and Evans, Philip M and Bamber, Jeffrey C and Bedford, James L},
	title = {Recent developments in non-coplanar radiotherapy},
	journal = {The British Journal of Radiology},
	volume = {92},
	number = {1097},
	pages = {20180908},
	year = {2019},
	doi = {10.1259/bjr.20180908},
	note ={PMID: 30694086}
}

@article{kingma2014adam,
	title={Adam: A method for stochastic optimization},
	author={Kingma, Diederik P and Ba, Jimmy},
	journal={arXiv preprint arXiv:1412.6980},
	year={2014}
}

@article{villa_fast_2022,
	title = {Fast {Monte} {Carlo}-{Based} {Inverse} {Planning} for {Prostate} {Brachytherapy} by {Using} {Deep} {Learning}},
	volume = {6},
	issn = {2469-7311, 2469-7303},
	doi = {10.1109/TRPMS.2021.3060191},
	number = {2},
	urldate = {2022-02-04},
	journal = {IEEE Transactions on Radiation and Plasma Medical Sciences},
	author = {Villa, Mateo and Bert, Julien and Valeri, Antoine and Schick, Ulrike and Visvikis, Dimitris},
	month = feb,
	year = {2022},
	pages = {182--188},
}

@inproceedings{ronneberger2015u,
  title={U-net: Convolutional networks for biomedical image segmentation},
  author={Ronneberger, Olaf and Fischer, Philipp and Brox, Thomas},
  booktitle={International Conference on Medical image computing and computer-assisted intervention},
  pages={234--241},
  year={2015},
  organization={Springer}
}

@article{nguyen2019feasibility,
  title={A feasibility study for predicting optimal radiation therapy dose distributions of prostate cancer patients from patient anatomy using deep learning},
  author={Nguyen, Dan and Long, Troy and Jia, Xun and Lu, Weiguo and Gu, Xuejun and Iqbal, Zohaib and Jiang, Steve},
  journal={Scientific reports},
  volume={9},
  number={1},
  pages={1--10},
  year={2019},
  publisher={Nature Publishing Group}
}

@article{nguyen2018three,
  title={Three-dimensional radiotherapy dose prediction on head and neck cancer patients with a hierarchically densely connected u-net deep learning architecture},
  author={Nguyen, Dan and Jia, Xun and Sher, David and Lin, Mu-Han and Iqbal, Zohaib and Liu, Hui and Jiang, Steve},
  journal={arXiv preprint arXiv:1805.10397},
  year={2018}
}

@inproceedings{cisternas2015matrad,
  title={matRad-a multi-modality open source 3D treatment planning toolkit},
  author={Cisternas, Eduardo and Mairani, A and Ziegenhein, P and J{\"a}kel, O and Bangert, M},
  booktitle={World Congress on Medical Physics and Biomedical Engineering, June 7-12, 2015, Toronto, Canada},
  pages={1608--1611},
  year={2015},
  organization={Springer}
}

@article{craft2014shared,
  title={Shared data for intensity modulated radiation therapy (IMRT) optimization research: the CORT dataset},
  author={Craft, David and Bangert, Mark and Long, Troy and Papp, D{\'a}vid and Unkelbach, Jan},
  journal={GigaScience},
  volume={3},
  number={1},
  pages={2047--217X},
  year={2014},
  publisher={Oxford University Press}
}

@dataset{test,
  author = {Wee, L. and Dekker, A.},
  title = {Data from Head-Neck-Radiomics-HN1, The Cancer Imaging Archive, https://wiki.cancerimagingarchive.net},
  year = {2019},
  doi = {https://doi.org/10.7937/tcia.2019.8kap372n}
}

@article{liu2021neuraldao,
  title={NeuralDAO: Incorporating neural network generated dose into direct aperture optimization for end-to-end IMRT planning},
  author={Liu, Cong and Ni, Xinye and Jin, Xiance and Si, Wen},
  journal={Medical Physics},
  year={2021},
  publisher={Wiley Online Library}
}

@article{bortfeld2006imrt,
  title={IMRT: a review and preview},
  author={Bortfeld, Thomas},
  journal={Physics in Medicine \& Biology},
  volume={51},
  number={13},
  pages={R363},
  year={2006},
  publisher={IOP Publishing}
}

@article{fan2019automatic,
  title={Automatic treatment planning based on three-dimensional dose distribution predicted from deep learning technique},
  author={Fan, Jiawei and Wang, Jiazhou and Chen, Zhi and Hu, Chaosu and Zhang, Zhen and Hu, Weigang},
  journal={Medical physics},
  volume={46},
  number={1},
  pages={370--381},
  year={2019},
  publisher={Wiley Online Library}
}

@article{hasenbalg2007collapsed,
  title={Collapsed cone convolution and analytical anisotropic algorithm dose calculations compared to VMC++ Monte Carlo simulations in clinical cases},
  author={Hasenbalg, Federico and Neuenschwander, H and Mini, Roberto and Born, Ernst Johann},
  journal={Physics in Medicine \& Biology},
  volume={52},
  number={13},
  pages={3679},
  year={2007},
  publisher={IOP Publishing}
}

@book{jenkins2012monte,
  title={Monte Carlo transport of electrons and photons},
  author={Jenkins, Theodore M and Nelson, Walter R and Rindi, Alessandro},
  volume={38},
  year={2012},
  publisher={Springer Science \& Business Media}
}

@article{jelen2005finite,
  title={A finite size pencil beam for IMRT dose optimization},
  author={Jele{\'n}, U and S{\"o}hn, M and Alber, M},
  journal={Physics in Medicine \& Biology},
  volume={50},
  number={8},
  pages={1747},
  year={2005},
  publisher={IOP Publishing}
}

\end{document}